\documentclass[conference,letterpaper]{IEEEtran}
\IEEEoverridecommandlockouts
\usepackage{cite}
\usepackage{amsmath,amssymb,amsfonts}
\usepackage{algorithmic}
\usepackage{graphicx}
\usepackage{pgf}
\usepackage{textcomp}
\usepackage[dvipsnames]{xcolor}
\usepackage{multirow}
\usepackage[caption=false,font=footnotesize]{subfig}

\begin{document}

\setlength{\columnsep}{0.24in} % IEEE minimum gutter

\title{Studying the Effect of Schedule Preemption on Dynamic Task Graph Scheduling
\thanks{This work was supported in part by the Army Research Laboratory under Cooperative Agreement W911NF-17-2-0196.}
}

\title{Studying the Effect of Schedule Preemption on Dynamic Task Graph Scheduling}

\author{
\IEEEauthorblockN{Mohammadali Khodabandehlou\IEEEauthorrefmark{1}, Jared Coleman\IEEEauthorrefmark{2}, Niranjan Suri\IEEEauthorrefmark{3}, Bhaskar Krishnamachari\IEEEauthorrefmark{1}}
\IEEEauthorblockA{\IEEEauthorrefmark{1}Dept. of Electrical and Computer Engineering, University of Southern California, Los Angeles, USA\\
\IEEEauthorrefmark{2}Dept. of Computer Science, Loyola Marymount University, Los Angeles, USA\\
\IEEEauthorrefmark{3}US Army Research Laboratory, Maryland, USA}
}

\maketitle

\maketitle

\begin{abstract}
Dynamic scheduling of task graphs is often addressed without revisiting prior task allocations, with a primary focus on minimizing makespan. We study \textit{controlled schedule preemption}, introducing the \textbf{Last-K Preemption} model, which selectively reschedules recent task graphs while preserving earlier allocations. Using synthetic, RIoTBench, WFCommons, and adversarial workloads, we compare preemptive, non-preemptive, and partial-preemptive strategies across makespan, fairness, utilization, and runtime. Results show moderate preemption can match most makespan/utilization gains of full preemption while maintaining fairness and low overhead.
\end{abstract}

% === IEEE Copyright + DOI block (for arXiv upload) ===
\begin{center}
\footnotesize
© 2025 IEEE. Personal use of this material is permitted. Permission from IEEE 
must be obtained for all other uses, in any current or future media, including 
reprinting/republishing this material for advertising or promotional purposes, 
creating new collective works, for resale or redistribution to servers or lists, 
or reuse of any copyrighted component of this work in other works.

This is the accepted version of the paper:
Mohammadali Khodabandehlou, Jared Coleman, Niranjan Suri, and Bhaskar Krishnamachari, “Studying the Effect of Schedule Preemption on Dynamic Task Graph Scheduling,” in Proc. IEEE Military Communications Conference (MILCOM) 2025, October 2025. DOI: 10.1109/MILCOM64451.2025.11310446.
\end{center}

\section{Introduction}
Scheduling computational tasks on distributed, heterogeneous networks is NP-hard \cite{garey1979computers, ullman1975np}, traditionally studied in static settings with a single Directed Acyclic Graph (DAG) and full workload knowledge \cite{kwok1999static, casavant2002taxonomy}. Real-world IoT and mission-critical systems require \textit{dynamic} scheduling where DAGs arrive unpredictably, and decisions must be made online.

Two paradigms exist: (1) \textbf{Preemptive}; rescheduling all pending tasks when a new DAG arrives; (2) \textbf{Non-preemptive}; preserving prior allocations and scheduling new DAGs on remaining resources. While preemption offers flexibility, it can harm fairness or incur overhead. We propose \textbf{Last-K Preemptive} scheduling: selectively rescheduling only the most recent $K$ DAGs to balance adaptability and stability.

Our contributions:
\begin{itemize}
    \item Define a richer evaluation suite that captures both performance and fairness.
    \item Compare preemptive, non-preemptive, and partial preemptive variants of classic heuristics.
    \item Demonstrate that moderate preemption yields near-optimal makespan/utilization without major fairness loss.
\end{itemize}

% \begin{itemize}\setlength\itemsep{0pt}\setlength\topsep{2pt}
%     \item Define a richer evaluation suite that captures both performance and fairness.
%     \item Compare preemptive, non-preemptive, and partial preemptive variants of classic heuristics.
%     \item Show that moderate preemption yields near-optimal makespan/utilization with low fairness cost.
% \end{itemize}

\section{Problem Definition \& Motivation}
We consider a new problem in which, instead of a single task graph, we have a set of task graphs that arrive over time. Specifically, let \(\{G_1, G_2, \ldots, G_K\}\) be a collection of task graphs, where each task graph is defined as \(G_i = (T_i, D_i)\) and arrives at time \(a_i \ge 0\). For each task \(t \in T_i\), the compute cost is given by \(c(t) \in \mathbb{R}^+\), and for each dependency \((t,t') \in D_i\), the data size is \(c(t,t') \in \mathbb{R}^+\). The compute node network is described by \(N = (V, E)\), a complete undirected graph where each node \(v \in V\) has a compute speed \(s(v) \in \mathbb{R}^+\), and each edge \((v,v') \in E\) has a communication strength \(s(v,v') \in \mathbb{R}^+\). In the related machines model~\cite{graham1969bounds}, the execution time of a task \(t \in T_i\) on a node \(v\) is given by \(\frac{c(t)}{s(v)}\), and the communication time between two tasks with dependency \((t,t')\), where \(t\) is executed on node \(v\) and \(t'\) on node \(v'\), is \(\frac{c(t,t')}{s(v,v')}\).

A valid schedule is an assignment of each task \(t \in T_i\) to a node \(v \in V\) along with a start time \(r(t)\) and a finish time \(e(t)\), such that the following conditions hold:
\begin{itemize}
 \item All tasks must be scheduled. For every task \(t \in T_i\) (for all \(i\)), there exists a scheduled assignment \(S_{A,N,G}(t) = (v, r, e)\) with \(v \in V\) and \(0 \le r \le e\).
    \item Execution times are valid. For every task \(t \in T_i\), if \(S_{A,N,G}(t) = (v, r, e)\) then the duration satisfies
    \[
    e - r = \frac{c(t)}{s(v)}.
    \]
    \item For any two distinct tasks \(t, t' \in T_i\) (or across different task graphs), if \(S_{A,N,G}(t) = (v, r, e)\) and \(S_{A,N,G}(t') = (v, r', e')\), then their execution intervals do not overlap, i.e., 
    \[
    e \le r' \quad \text{or} \quad e' \le r.
    \]
    \item For any task \(t \in T_i\), its execution cannot begin before the arrival time \(a_i\) of its corresponding task graph \(G_i\), i.e., 
    \[
    r(t) \ge a_i
    \].
    \item For every dependency \((t,t') \in D_i\), if task \(t\) is scheduled on node \(v\) and task \(t'\) on node \(v'\), then \(t'\) may only start after \(t\) has completed and its output has been communicated, i.e.,
    \[
    e(t) + \frac{c(t,t')}{s(v,v')} \le r(t')
    \]
\end{itemize}

The objective is to determine a schedule for all task graphs in \(\{G_1, G_2, \ldots, G_K\}\) on the network \(N\) that minimizes the overall makespan, defined as the maximum completion time over all tasks.

\begin{figure}[t]
    \centering
    \subfloat[P-HEFT Scheduler]{%
        \includegraphics[width=0.86\columnwidth]{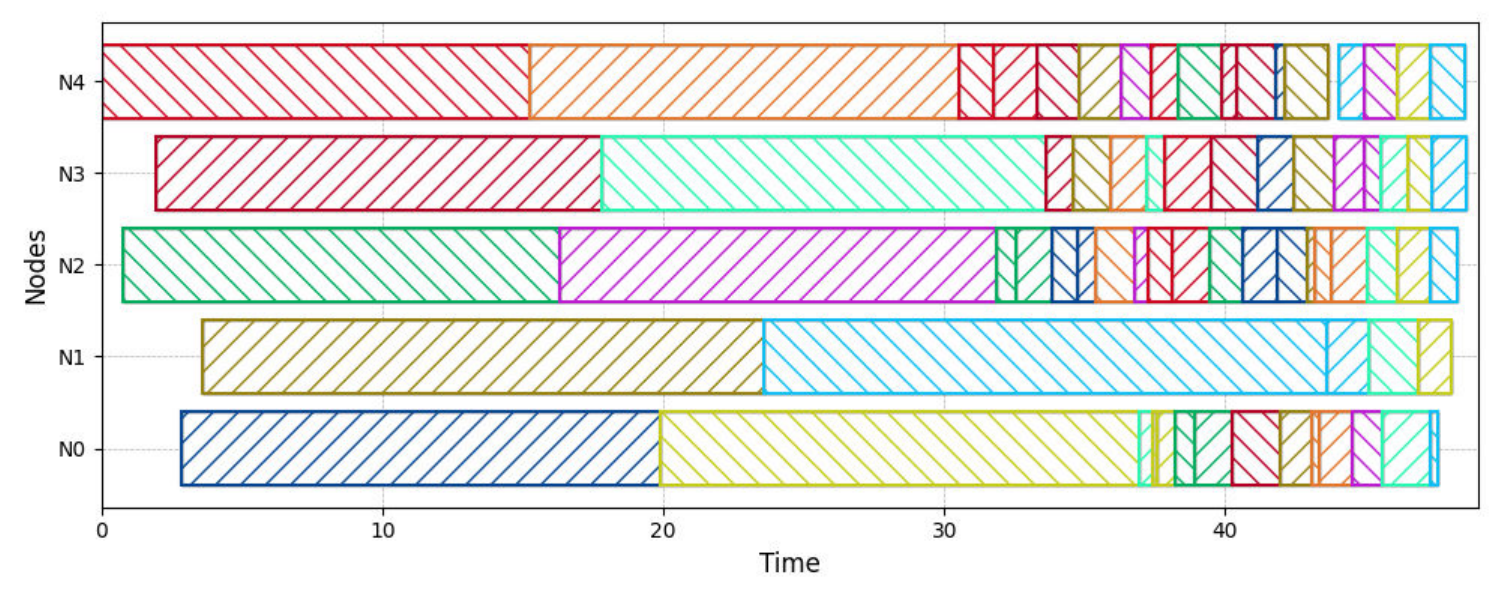}}\par
    \subfloat[5P-HEFT Scheduler]{%
        \includegraphics[width=0.86\columnwidth]{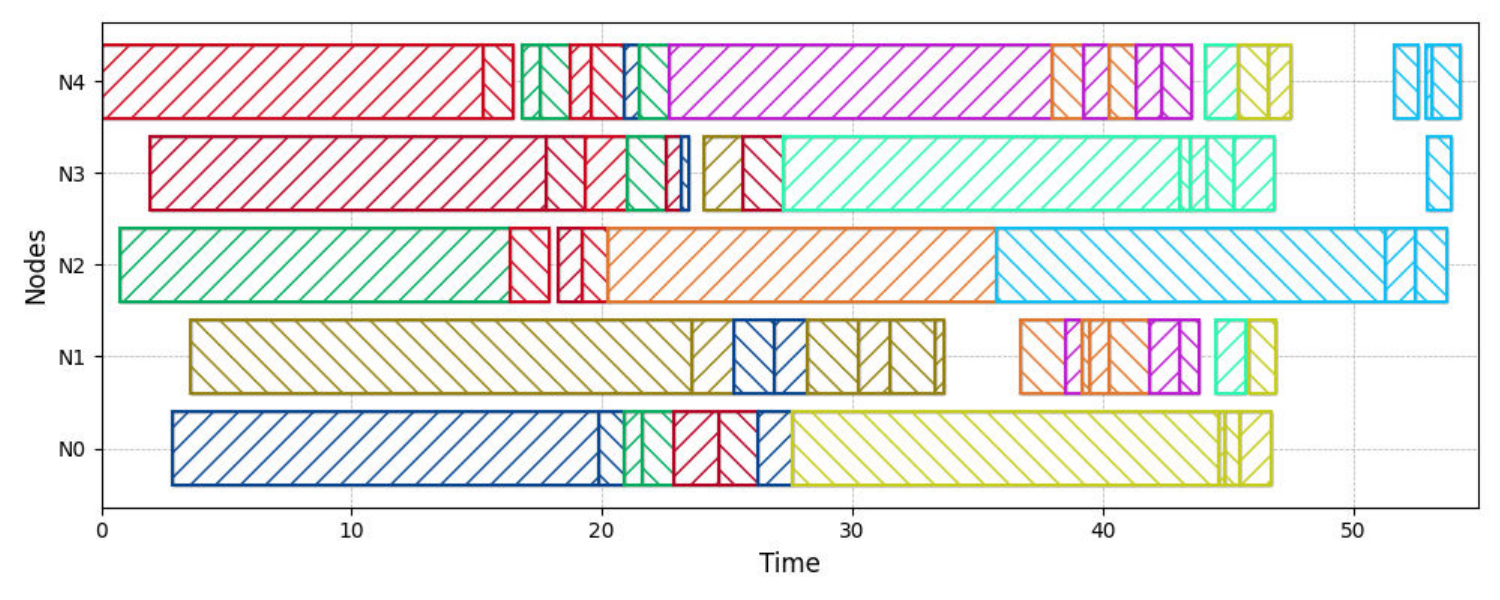}}\par
    \subfloat[NP-HEFT Scheduler]{%
        \includegraphics[width=0.86\columnwidth]{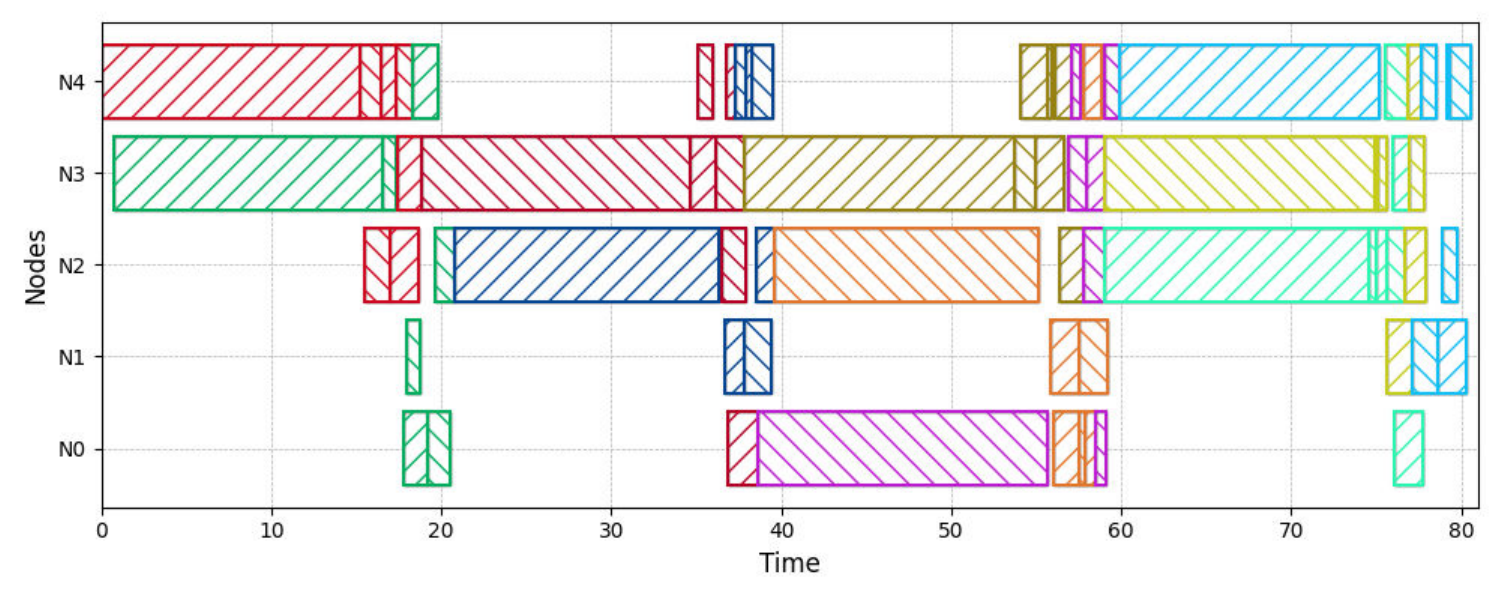}}
    \caption{Large leading tasks being blocked by small tasks from previous task graphs}
    \label{example3}
\end{figure}

This problem is motivated by dynamic, heterogeneous environments such as IoBT, mission-critical systems, and distributed workflows, where task graphs arrive unpredictably over time. Prior work, including our preliminary study presented in \cite{khodabandehlou2025scheduling}, has demonstrated that dynamic scheduling without reconsidering prior allocations can lead to inefficiencies in both makespan, while uncontrolled rescheduling can adversely affect fairness. For instance, when a task graph has a large root followed by many small tasks, a non-preemptive scheduler cannot displace small tasks from earlier graphs, which can greatly increase makespan and reduce utilization. In Figure~\ref{example3}.c, the non-preemptive scheduler's makespan is much larger than the preemptive one due to this limitation. However, as seen in Figure~\ref{example3}.a, the preemptive scheduler, while faster overall, delays small tasks, leading to poor fairness.

In this work, we expand on those findings by formalizing the problem, introducing controlled preemption policies, and evaluating them across diverse workload classes.

\section{Related Work}
List scheduling heuristics such as HEFT and CPOP \cite{topcuoglu2002performance} dominate static DAG scheduling, with metaheuristics offering improvements at higher computational cost \cite{zomaya1999genetic}. Dynamic and online variants \cite{arabnejad2012fairness} adapt to workload arrivals but rarely revisit earlier allocations. Recent work integrates deep reinforcement learning and graph neural networks for adaptability \cite{yanggraph, gu2025deep}.

Runtime frameworks such as StarPU \cite{augonnet2009starpu} and PaRSEC \cite{bosilca2013parsec} target high-performance computing (HPC) and supercomputing environments, operating on fine grained task graphs representing kernels, matrix factorizations, or other small compute units. These systems assume tightly coupled clusters with low-latency interconnects, and focus on optimizing execution at the kernel level rather than handling dynamic, unpredictable task graph arrivals.

In contrast, our work addresses coarser grained tasks at the container or workflow level, executing on heterogeneous, potentially resource constrained networks. Jupiter~\cite{ghosh2021jupiter} targets container-level DAGs in distributed/edge settings but does not study schedule-level preemption.. To the best of our knowledge, this work is the first to systematically study controlled preemption strategies for dynamic, heterogeneous DAG scheduling, evaluating trade-offs in makespan, fairness, utilization, and runtime.

% requires: \usepackage[caption=false,font=footnotesize]{subfig}

\section{Scheduling Paradigms}

\subsection{Preemptive Scheduler}
\begin{figure}[t]
  \centering
  \includegraphics[width=\linewidth]{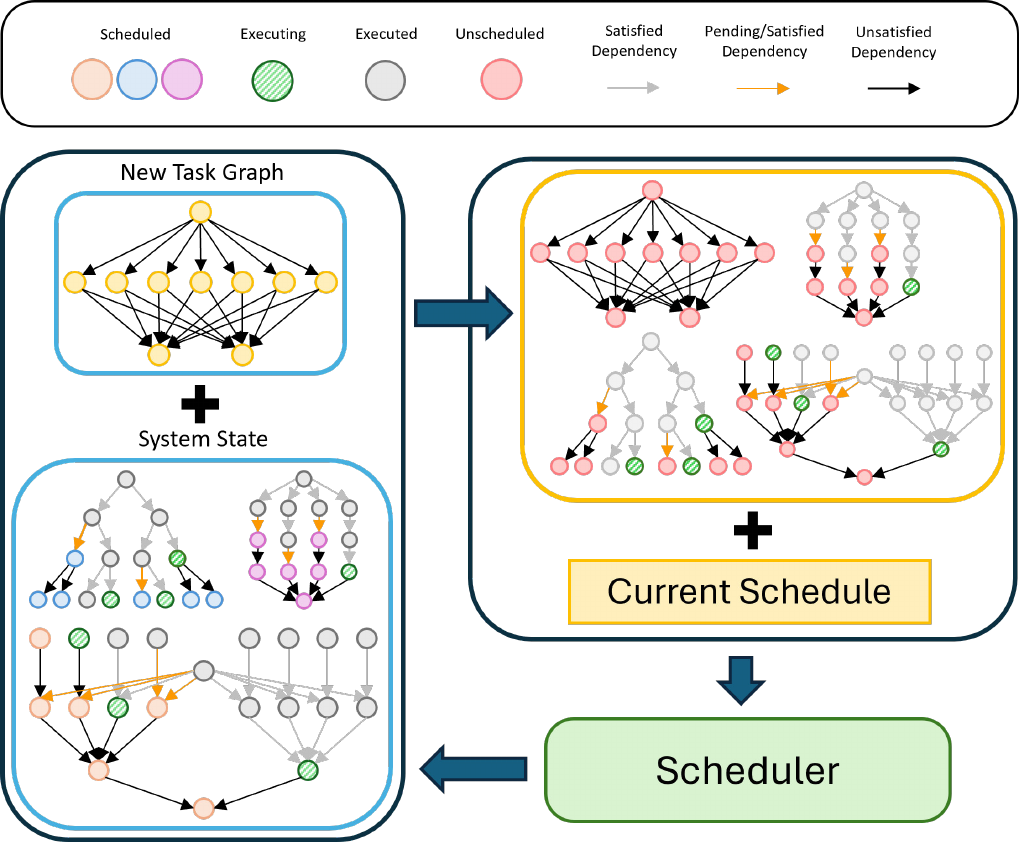}
  \caption[Preemptive Scheduler]{%
    Preemptive Scheduler. Having different colors for \textit{Scheduled} tasks means that they belong to different task graphs.
  }
  \label{preemptive_scheduler}
\end{figure}

When a new task graph arrives, we merge it with all pending tasks into a single graph with multiple connected components (Figure~\ref{preemptive_scheduler}). All \textbf{Scheduled} tasks are reverted to \textbf{Unscheduled}, and the updated graph is resubmitted to the scheduler for a new allocation. 
% This work does not model dependency satisfaction or bandwidth wasted by rescheduling, though incorporating these factors could further reduce makespan and improve accuracy. Their omission does not affect the generality of our approach.

\subsection{Non-Preemptive Scheduler}
In the non-preemptive approach, tasks marked as \textbf{Scheduled} remain fixed, and the new task graph is placed only on available resources. Prior tasks continue execution and data transfer without modification. Similar strategies appear in Dynamic Workflow Scheduling (DWS) and Job Shop Scheduling (JSSP), though typically without modeling network cost or heterogeneity \cite{yanggraph, huang2022cost, jadeja2012cloud}. To our knowledge, this is the first direct comparison of preemptive and non-preemptive scheduling in such settings.

\subsection{Partially Preemptive Scheduler}
To balance flexibility and stability, we introduce \textbf{Last-K Preemptive} scheduling, which reschedules tasks only from the most recent $K$ task graphs. While not necessarily optimal, this approach offers the adaptability of preemption while avoiding its fairness and runtime penalties, and can be integrated into existing schedulers with minimal complexity. In Figure~\ref{example3}.b, the last-5 preemptive HEFT variant (5P-HEFT) achieves a much smaller makespan than NP-HEFT (Figure~\ref{example3}.c) without the fairness issues of P-HEFT (Figure~\ref{example3}.a).

\section{Metrics}
Relying solely on makespan overlooks key trade-offs between performance and fairness. We evaluate schedulers using the following metrics.

\subsection{Total Makespan}
The total makespan, a common performance metric, measures time from the arrival of the first task graph to the completion of the last:
\[
\max_{i,\, t \in T_i} e(t)
\]
where $T_i$ is the set of tasks in $G_i$ and $e(t)$ the finish time of task $t$. While useful for overall throughput, it ignores responsiveness to individual graphs.

\subsection{Mean Makespan}
Mean makespan captures per-graph responsiveness:
\[
\frac{1}{K}\sum_{i=1}^K\left(\max_{t \in T_i} e(t)-a_i\right)
\]
where $K$ is the number of graphs, $a_i$ the arrival time of $G_i$. It reflects both performance and fairness by indicating the average time each graph remains in the system, making it relevant in deadline-sensitive settings.

\subsection{Mean Flowtime}
Mean flowtime measures how compactly tasks of a graph are executed:
\[
\frac{1}{K}\sum_{i=1}^K\left(\max_{t \in T_i} e(t) - \min_{t' \in T_i} r(t')\right)
\]
where $r(t)$ is the start time of task $t$. As a fairness metric, lower values imply contiguous execution, important in scenarios like stream processing or iterative ML, where spread-out execution increases staleness and overhead.

\subsection{Node Utilization}
Node utilization measures effective use of compute resources:
\[
u(v) = \frac{\sum_{\substack{i,t \in T_i \\ S_A(t) = v}} \frac{c(t)}{s(v)}}{\max_{i,\, t \in T_i} e(t)}
\]
where $S_A(t)$ is the assigned node, $c(t)$ the computation cost, and $s(v)$ the node speed. Higher utilization indicates better resource leverage.

\subsection{Runtime}
Runtime is the time the scheduler takes to compute a valid schedule, including priority evaluation and resource assignment. While not affecting final schedule quality, high runtime can limit applicability in real-time or high-arrival-rate environments.

\begin{figure}[t]
    \centering
    \subfloat[Synthetic]{%
        \resizebox{0.9\columnwidth}{!}{\input{images/makespan-synthetic.pgf}}}\par
    \subfloat[RIoTBench]{%
         \resizebox{0.9\columnwidth}{!}{\input{images/makespan-riotbench.pgf}}}\par
    \subfloat[WFCommons]{%
         \resizebox{0.9\columnwidth}{!}{\input{images/makespan-wfcommons.pgf}}}\par
    \caption{Normalized Makespan}
    \label{makespan}
\end{figure}

\section{Evaluation}
All experiments were implemented in Python within a SAGA simulation~\cite{coleman2024comparing} environment, which provides reference schedulers including HEFT~\cite{topcuoglu1999task}, CPOP~\cite{topcuoglu1999task}, Min-Min~\cite{braun2001comparison}, Max-Min~\cite{braun2001comparison}, and Random. We evaluate using both synthetic and real-world workflows from RIoTBench~\cite{shukla2017riotbench} and WFCommons~\cite{coleman2023automated}.

\subsection{Synthetic Task Graphs}
We generate 100 graphs evenly split among four structures: \textbf{Out Tree}, \textbf{In Tree}, \textbf{Fork Join}, and \textbf{Chain}. Task and edge weights follow a 5-component truncated Gaussian mixture; node speeds and link rates come from single truncated Gaussians. This setup models varied dependency topologies common in parallel and streaming workloads.

\begin{figure}[t]
    \centering
    \subfloat[Synthetic]{%
        \resizebox{0.86\columnwidth}{!}{\input{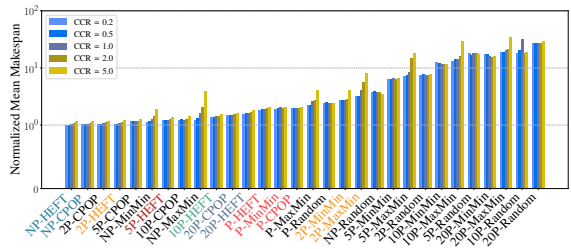}}}\par
    \subfloat[RIoTBench]{%
        \resizebox{0.86\columnwidth}{!}{\input{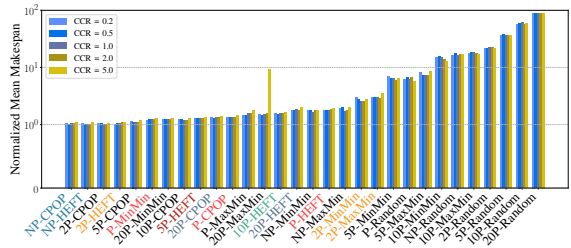}}}\par
    \subfloat[WFCommons]{%
        \resizebox{0.86\columnwidth}{!}{\input{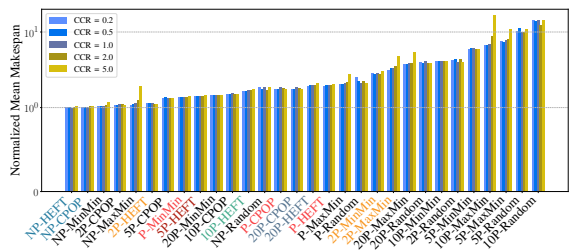}}}
    \caption{Normalized Mean Makespan}
    \label{mean_makespan}
\end{figure}

\begin{figure}[t]
    \centering
    \subfloat[Synthetic]{%
        \resizebox{0.86\columnwidth}{!}{\input{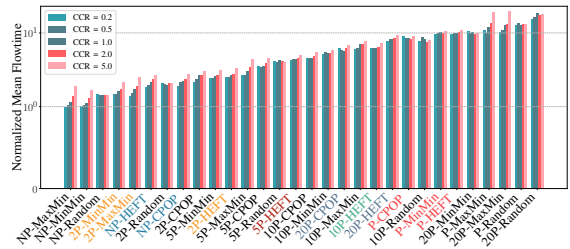}}}\par
    \subfloat[RIoTBench]{%
        \resizebox{0.86\columnwidth}{!}{\input{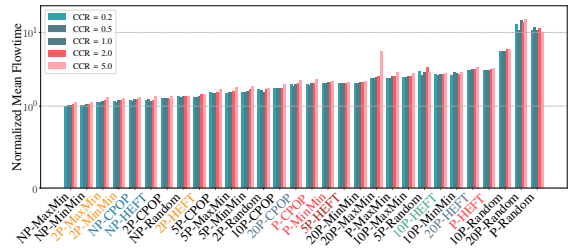}}}\par
    \subfloat[WFCommons]{%
        \resizebox{0.86\columnwidth}{!}{\input{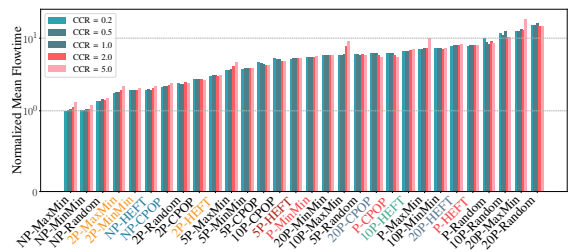}}}
    \caption{Normalized Mean Flowtime}
    \label{mean_flowtime}
\end{figure}

\subsection{RIoTBench Workflows}
From RIoTBench, we use \textbf{ETL}, \textbf{Predict}, \textbf{Stats}, and \textbf{Train} DAGs, instantiating 100 graphs with equal type probability. These preserve original topologies and operator structures, introducing greater heterogeneity and imbalance than synthetic graphs, and reflecting realistic IoT processing pipelines.

\subsection{WFCommons Workflows}
We select nine scientific workflows: \textbf{Epigenomics}, \textbf{Montage}, \textbf{Cycles}, \textbf{Seismology}, \textbf{SoyKB}, \textbf{SRA Search}, \textbf{Genome}, \textbf{Blast}, and \textbf{BWA} for 50 total graphs, evenly distributed by type. DAGs retain original dependencies and task parameters, featuring long critical paths and complex communication, making them well-suited for testing scalability, fairness, and stability in large-scale distributed settings.

\subsection{Adversarial Instance}
To probe worst-case behavior, we evaluate on adversarial instances inspired by Figure~\ref{example3}. Each instance is an \textbf{Out Tree} with a large-computation root followed by many shallow, lightweight successors. This forces the root to finish before any successor can run, creating a bottleneck. We set the Communication-to-Computation Ratio (CCR) to 0.2, making communication costs negligible and encouraging schedulers to spread successors across processors—often causing underutilization and idle gaps (Figure~\ref{example3}).

\section{Results}
We refer to preemptive, non-preemptive, and partially preemptive schedulers as P-NAME, NP-NAME, and KP-NAME, respectively. For example, P-MaxMin, NP-HEFT, and 5P-CPOP denote preemptive Max-Min, non-preemptive HEFT, and Last-5 partially preemptive CPOP.

\begin{figure}[t]
    \centering
    \subfloat[Synthetic]{%
        \resizebox{0.86\columnwidth}{!}{\input{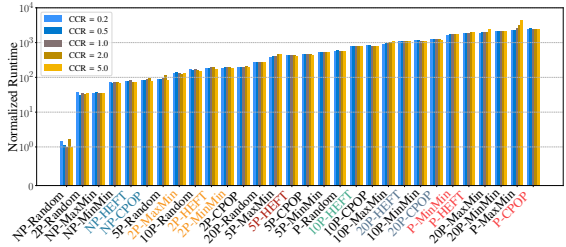}}}\par
    \subfloat[RIoTBench]{%
        \resizebox{0.86\columnwidth}{!}{\input{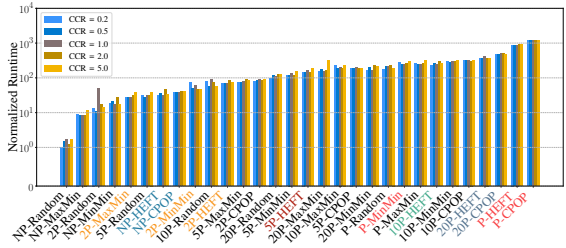}}}\par
    \subfloat[WFCommons]{%
        \resizebox{0.86\columnwidth}{!}{\input{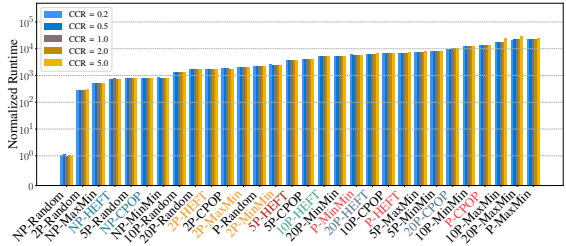}}}
    \caption{Normalized Runtime}
    \label{runtime}
\end{figure}

\subsection{Total Makespan}
Across all datasets (Figure~\ref{makespan}), preemptive schedulers generally achieve smaller makespans, as they can move tasks to reduce completion time. However, the gap between the best preemptive and non-preemptive schedulers is smaller than expected in most cases, likely due to high utilization. When utilization is already high, the room for improvement from preemption is limited.  
In the adversarial instance (Figure~\ref{adversarial}.a), this gap widens dramatically: \textbf{NP-HEFT}’s makespan is 1.6$\times$ that of \textbf{P-HEFT}, showing how non-preemptive schedulers struggle with blocking tasks. Partially preemptive schedulers such as \textbf{20P-HEFT} and \textbf{10P-HEFT} perform nearly as well as \textbf{P-HEFT} while avoiding excessive preemption.

\subsection{Mean Makespan}
Mean makespan (Figure~\ref{mean_makespan}) measures the average time from task graph arrival to completion. On regular workloads, non-preemptive schedulers lead, as they avoid delaying tasks through rescheduling. Partially preemptive schedulers follow closely, offering a balance between responsiveness and flexibility.  
In the adversarial case (Figure~\ref{adversarial}.b), partially preemptive schedulers achieve the lowest mean makespan. Fully preemptive schedulers can indefinitely delay some tasks to minimize total makespan, while non-preemptive ones are blocked by small tasks, widening the gap between arrival and completion.

\begin{figure}[t]
    \centering
    \subfloat[Synthetic]{%
        \resizebox{0.86\columnwidth}{!}{\input{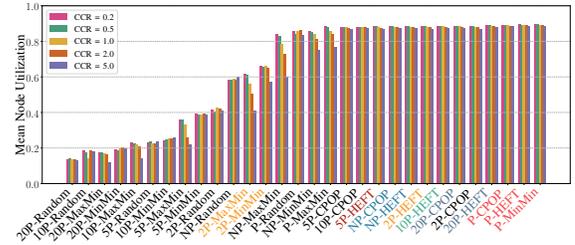}}}\par
    \subfloat[RIoTBench]{%
        \resizebox{0.86\columnwidth}{!}{\input{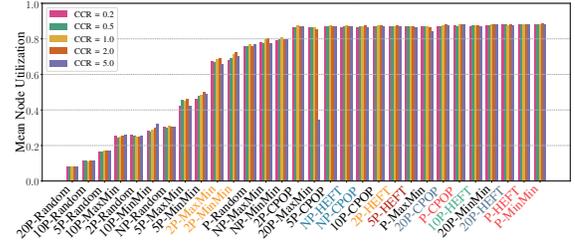}}}\par
    \subfloat[WFCommons]{%
        \resizebox{0.86\columnwidth}{!}{\input{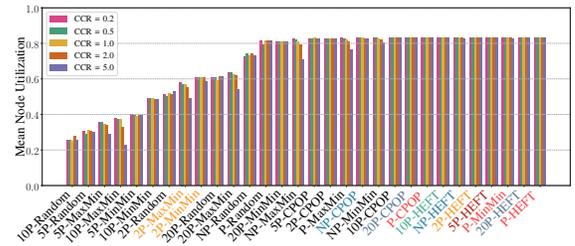}}}
    \caption{Utilization}
    \label{utilization}
\end{figure}

\subsection{Mean Flowtime}
Mean flowtime (Figure~\ref{mean_flowtime}) reflects fairness by measuring how compactly tasks of a graph are scheduled. Non-preemptive schedulers consistently produce the smallest flowtimes, a trend that holds even at higher arrival rates since flowtime is independent of graph arrival time. Partially preemptive schedulers such as \textbf{2P-MaxMin} and \textbf{2P-MinMin} maintain comparably low values.  
Under the adversarial workload (Figure~\ref{adversarial}.c), \textbf{5P-HEFT} and \textbf{20P-CPOP} match or exceed the fairness of \textbf{NP-HEFT} and \textbf{NP-CPOP}, while also keeping the makespan small.

\subsection{Runtime}
\begin{figure}[t]
    \centering
    \subfloat[Makespan]{%
        \resizebox{!}{1.1in}{\input{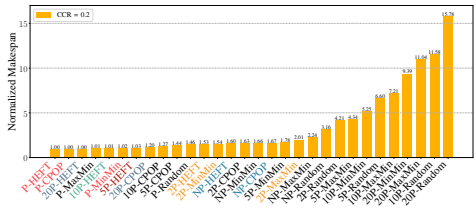}}}\par
    \subfloat[Normalized Mean Makespan]{%
        \resizebox{!}{1.1in}{\input{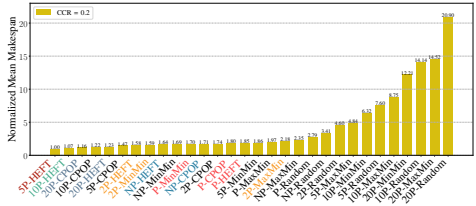}}}\par
    \subfloat[Normalized Mean Flowtime]{%
        \resizebox{!}{1.1in}{\input{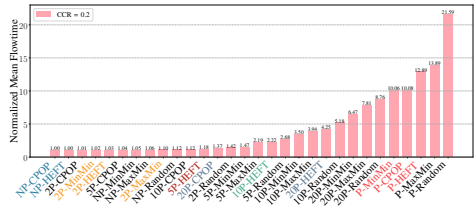}}}\par
    \subfloat[Normalized Runtime]{%
        \resizebox{!}{1.1in}{\input{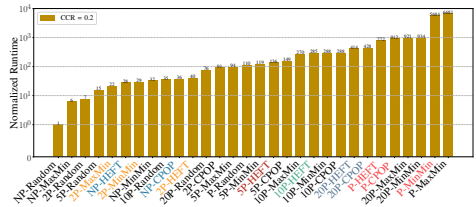}}}\par
    \subfloat[Utilization]{%
        \resizebox{!}{1.1in}{\input{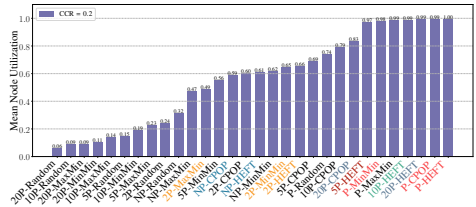}}}
    \caption{Adversarial Instance}
    \label{adversarial}
\end{figure}

Runtime here is the total time from the first task graph’s arrival to scheduling the last one (Figure~\ref{runtime}). Non-preemptive schedulers are fastest, as they only place new tasks on the remaining resources. Partially preemptive schedulers with low $K$, such as \textbf{2P-MaxMin}, \textbf{2P-MinMin}, and \textbf{2P-HEFT} follow, while fully preemptive schedulers (\textbf{P-HEFT}, \textbf{P-CPOP}) are slowest.  
The same trend appears in adversarial runs (Figure~\ref{adversarial}.d): \textbf{NP-HEFT} is fastest, and \textbf{5P-HEFT} achieves a runtime close to it, showing that limited preemption can preserve scheduling speed.

\subsection{Utilization}
Higher preemption generally increases utilization (Figure~\ref{utilization}). Fully preemptive schedulers (\textbf{P-HEFT}, \textbf{P-MinMin}, \textbf{P-CPOP}) achieve the best results. In synthetic and RIoTBench workloads, \textbf{MaxMin} and \textbf{MinMin} show poor utilization, while WFCommons reduces this gap, highlighting the importance of testing across diverse workloads. Higher CCR values tend to reduce utilization, as communication costs discourage task distribution.  
In the adversarial workload (Figure~\ref{adversarial}.e), utilization improves sharply from \textbf{5P-HEFT}, with many partially preemptive schedulers reaching levels close to the fully preemptive upper bound.

\section{Conclusion}
Controlled schedule preemption enables dynamic schedulers to trade off makespan, utilization, fairness, and overhead. Last-K Preemptive scheduling offers a balanced approach, matching much of the performance of full preemption while avoiding its fairness penalties and runtime costs.

\section*{Acknowledgments}
This work was supported in part by ARL under Cooperative Agreement W911NF-17-2-0196.

\bibliographystyle{IEEEtran}
\bibliography{references}

\end{document}